\documentstyle[prl,aps,epsf] {revtex}

\begin{document}
\title{Testing quantum correlations in a confined
atomic cloud by scattering fast atoms: Direct and time
reversed processes}
\author{A.B. Kuklov$^1$ and B.V. Svistunov$^2$}
\address{$^1$ Department of Engineering Science and Physics,
The College of  Staten Island, CUNY,
     Staten Island, NY 10314}
\address{$^2$ Russian Research Center ``Kurchatov Institute",
123182 Moscow, Russia}

\date{\today}
\maketitle

\begin{abstract}
We suggest measuring  the one-particle density matrix
of a trapped ultracold atomic cloud by scattering
fast atoms in a pure momentum state off the cloud.  
The lowest-order probability for the process,
resulting in a pair of outcoming fast
atoms for each incoming one, as well as of its time
reversed counterpart, turns out to be given by 
the Fourier transform of the density matrix. Accordingly,
important information about
quantum correlations can be deduced directly from
the differential scattering cross-section of these
processes. Several
most interesting cases of scattering - from a single 
condensate containing a vortex, and from a split 
condensate characterized by some phase difference -
are discussed. 
\\

\noindent PACS numbers: 03.75.Fi, 05.30.Jp, 32.80.Pj, 67.90.+z
\end{abstract}
\vskip0.5 cm

Achievements of collective quantum states 
in confined clouds of alkaline atoms \cite{BEC} 
and in atomic hydrogen \cite{HYDRO} make possible studying
quantum coherent properties of these systems as well as the
revealing fundamental kinetic processes leading to the 
formation of the coherence.
The coherence of the condensate has been   
demonstrated in 
the atomic interference  fringes \cite{COHER}. Recently 
Bragg spectroscopy has been successfully employed to establish
that the coherence length of a confined condensate is of the
order of the cloud size \cite{BRAGG}. 

The mechanism of formation of quantum correlations is 
a matter of great attention and controversy. The emergence of
the condensate 
 is associated with 
the formation of  off-diagonal long-range order
(ODLRO) \cite{ODLRO}. A primary object
displaying such an order is the one-particle density matrix
(OPDM) $\rho ({\bf x}_1,{\bf x}_2)$. Typical 
distances over which these correlations become important
are comparable with the interatomic spacing $r_a$. Consequently,
an ``early" detection of such emerging correlations is very difficult
to achieve by light with the wavelength 
$\lambda \gg r_a$ usually employed for probing the cloud density. 
Information about short-range density correlations 
(at distances $ < r_a$) can, in principle, be obtained
from the absorption of detuned resonant light \cite{ABSORB}.
The change in local $m$-body density correlations 
(the so-called $m!$-effect \cite{M!}) can be seen by measuring 
recombination rates, and this has been already done experimentally 
for the equilibrium case \cite {Burt}. However, measurements of
the correlation length $r_c$ (which is far less than
$\lambda $)  of the forming ODLRO seem very
unlikely to be achieved by these methods. 
Thus it is tempting to have a tool which could make possible seeing
the OPDM directly without fundamental limitations on the accessible distances. 

In this paper, we suggest a method of detecting the OPDM which relies 
on the scattering of coherent atoms off the atomic cloud.  

The methods of scattering neutrons \cite{NEUT1,NEUT2}
and He atoms \cite{ATOMS} off liquid He are well known. For an
incident neutron (or atom) which has a relatively small kinetic
energy if compared with a typical potential energy of atoms
in the target, a typical scattering event results in the creation 
of collective excitations. In contrast,
an incident fast neutron (atom) ejects an almost free fast 
atom from the liquid. While the former scenario has practically
the same features in both helium and weakly interacting 
Bose gases, the latter has very different consequences 
in the two systems. Indeed, the weakly interacting 
Bose gas is almost transparent for the fast atoms, provided
the cloud size $R$ traversed by them is less than the free-path 
length $l=1/na^2$, where $n$ and $a$ stand for the density and 
the scattering length, respectively.
We, however, note that
this condition is by no means fundamentally restrictive as long as 
$l$ is much larger than the interparticle distance $n^{-1/3}$.
In fact, the condition $l \gg n^{-1/3}$ is guaranteed 
by the smallness of the gas parameter $\xi =na^3$ 
which can be as small as $\xi \sim 10^{-5}$ in the atomic clouds. 
Thus, in a weakly interacting gas there is the opportunity
of measuring directly the outcome of a single scattering
event. We demonstrate that such a measurement, if 
accurate enough, yields direct information
on the OPDM in the Bose gas. This method can also
be applied to studying trapped Fermi gases. 
In contrast, this is impossible in principle in 
liquid helium where $l\approx n^{-1/3}$ which leads to
 dominance of the multiparticle excitations in the 
final-state channel. Accordingly, only some integral
of the population factors can be obtained from the scattering
cross-section \cite{NEUT1}. 

Here, we consider
the fast incoming atoms  
to be identical to the atoms of the cloud.  
A fast atom carrying
momentum ${\bf k}$ can transfer a
substantial part of its energy to one atom in the cloud,
so that in the final state there are
two fast outcoming atoms with momenta ${\bf k}_1$
and ${\bf k}_2 $. 
The quantum-mechanical probability
of this process in the lowest order turns out to be proportional 
to a Fourier transform of the OPDM. Let us call this process
{\it direct}. This process has been considered in Ref.\cite{CON}. 
There is another process which is connected by time reversal
with the {\it direct} one. Specifically, two fast incoming atoms
carrying the respective momenta ${\bf k}_1$ and ${\bf k}_2$
enter the cloud and experience stimulated (by the atoms of the cloud)
scattering, so that in the final state there is 
 only one fast 
atom with the momentum ${\bf k}$. We 
will call this process {\it reversed}. 

It should be mentioned that there is another process
of atomic scattering, which 
is analogous to the elastic scattering of light. 
Its probability depends on the cloud density
insensitive to $r_c$ \cite{ELASTIC}. 

An expression for the cross-section of
the {\it direct} and the {\it reversed} 
processes can be derived 
in the lowest order with respect to the two-body interaction
for the simplest case of spin-polarized bosons. 
The essential requirement is that the kinetic energy $k^2/2m$ 
 (we employ atomic units so that
$\hbar =1$) of a fast atom of mass $m$
is much larger than a
typical interaction energy per particle of the densest part of the
cloud. This typical energy is given by the chemical 
potential $\mu \approx 4\pi an/m$.
 Thus, one finds that $ka \gg \sqrt{8\pi na^3}\approx \xi^{1/2}$. 
In other words, the de Broglie wavelength of the incoming
atom must be shorter than the healing length $1/\sqrt{an}$ of the cloud.  
For the sake of simplicity, we also impose
an additional (less critical) requirement 
$ka \ll 1$ insuring that the $s$-wave scattering approximation is sufficient.
Combining these two requirements, one obtains the condition 

\begin{eqnarray}
\xi^{1/2}\ll ka \ll 1, 
\label{1}
\end{eqnarray}
\noindent 
which has a wide range of validity as long as the gas parameter
$\xi \ll 1$. 

One can represent the interaction Hamiltonian in the
traditional form

\begin{eqnarray}
H_{\mbox{\scriptsize int}} \, = \, {u_0 \over 2}\int d{\bf x} \,
\Psi^{\dagger}({\bf x})\Psi^{\dagger}({\bf x})
\Psi ({\bf x})\Psi ({\bf x}) \; ,  \quad u_0={4\pi a\over m} \; ,
\label{2}
\end{eqnarray}
\noindent
Under the condition (\ref{1}) the fast atoms behave almost
like free particles, and their interaction with the rest of the 
gas can be treated perturbatively. Hence, it is reasonable to subdivide 
the total field $\Psi$ into the low- and the high-energy parts, 
$\psi$ and $\psi '$, respectively:

\begin{eqnarray}
\Psi = \psi + \psi' \; , \quad \psi' =\sum_{\bf k} a_{\bf k} \,
{\rm e}^{i{\bf kx}} \; ,
\label{3}
\end{eqnarray}
where the incident and the scattering states
are described in terms of plane waves  
normalized to unit volume; $a_{\bf k}$ destroys the high-energy particle
with the momentum ${\bf k}$. 
Substitution
of Eq.~(\ref{3}) into Eq.~(\ref{2})
and selection of the terms that describe
the {\it direct} and the {\it reversed} processes 
 yields

\begin{eqnarray}
H'_{\mbox{\scriptsize int}} \, = \, u_0\sum_{{\bf k}_1,{\bf k}_2,{\bf k}}
a^{\dagger}_{{\bf k}_1} a^{\dagger}_{{\bf k}_2} a_{\bf k}
\psi_{\bf q} \, + \, \mbox{H.c.} \; ,
\label{4}
\end{eqnarray}
\noindent
where the Fourier component ~$\psi_{\bf q}= \int d{\bf x} \, \exp (-i{\bf qx})\psi ({\bf x})$~ of the field $\psi ({\bf x})$~ is introduced, and
~${\bf q}={\bf k}_1 + {\bf k}_2 - {\bf k}$~ should be interpreted
as the momentum transferred from (to) the cloud.

First, let us consider the {\it direct} process. Then, in the initial
state there are a fast atom with the momentum ${\bf k}$ and the cloud
of $N$ atoms. In the final
state there are two fast atoms with the momenta ${\bf k}_1,\; {\bf k}_2$
and the cloud of $N-1$ atoms. It is assumed
that two outcoming atoms can be identified \cite{PAIR}, so that
the momentum transfer ${\bf q}$ and the energy transfer $\omega
=(k^2_1+k^2_2 -k^2)/2m$ can
both be measured. Then, 
the corresponding double-differential cross-section for the direct 
process in the lowest order with respect to $H'_{\mbox{\scriptsize int}}$ 
is given by the Golden rule formula

\begin{eqnarray}
\displaystyle W({\bf q},\omega)=
{2 m u_0^2 \over k}\int {d{\bf k}_1\over (2\pi )^3} \,
\delta (\omega - \omega_{fi})
\int \! \! \int d{\bf x}_1 d{\bf x}_2\int_{-\infty}^{\infty} dt \,
{\rm e}^{i{\bf q}({\bf x}_1-{\bf x}_2)-i\omega t} \,
\rho ({\bf x}_1, t; {\bf x}_2, 0) \; .
\label{5}
\end{eqnarray}
\noindent
Here $\rho ({\bf x}_1, t_1; {\bf x}_2, t_2)= 
\langle \psi^{\dagger}({\bf x}_1, t_1)
\psi ({\bf x}_2, t_2)\rangle $ 
[Note that unlike the uniformity in time, 
~$\rho ({\bf x}_1, t_1; {\bf x}_2, t_2)=
\rho ({\bf x}_1, t_1-t_2; {\bf x}_2, 0)$,
the space uniformity cannot be, in general, assumed as long 
as a trapping potential exists]; $\omega_{fi}=
 [ q^2/2+ {\bf qk}+
 k^2_1- ({\bf q} + {\bf k}){\bf k_1} ]/m $
is the difference of the kinetic energies 
of the fast atoms in the final and initial states
where
the momentum of the second atom, ${\bf k}_2$,
is fixed by the relation 
${\bf k}_2= {\bf q}+{\bf k} - {\bf k}_1$. 

The integration over ${\bf k}_1$ in Eq.(\ref{5})
can be carried out explicitly. However, first
we notice that the requirement of large $k$ means that the values of 
$q$ and $\omega$, which are effectively selected by the correlator 
$\rho ({\bf x}_1, t; {\bf x}_2, 0)$ in  
Eq.~(\ref{5}), satisfy the conditions $q \ll k$ and
$|\omega | \ll  k^2 /m$. This immediately leads to the approximation
$\delta (\omega - \omega_{fi}) \approx m \, \delta (k_1^2 - {\bf k k}_1 )$,
which yields
 
\begin{eqnarray}
\displaystyle W({\bf q},\omega)=
4a^2 \int \! \! \int d{\bf x}_1 d{\bf x}_2\int_{-\infty}^{\infty} dt \,
{\rm e}^{i{\bf q}({\bf x}_1-{\bf x}_2)-i\omega t} \, 
\rho ({\bf x}_1, t; {\bf x}_2, 0) \; . 
\label{7}
\end{eqnarray}
\noindent
We thus see that $W({\bf q},\omega)$
is directly related to the dynamic correlator
$\rho ({\bf x}_1, t; {\bf x}_2, 0)$ which, contains
rather rich information about the system, including, for one thing,
 the elementary
excitation spectrum. Confining ourselves to the static correlations,
described by OPDM $\rho ({\bf x}_1,{\bf x}_2) = 
\rho ({\bf x}_1, 0; {\bf x}_2, 0)$, we arrive at the even simpler relation

\begin{eqnarray}
\displaystyle W({\bf q})=
8 \pi a^2 \int \! \! \int d{\bf x}_1 d{\bf x}_2 \, 
{\rm e}^{i{\bf q}({\bf x}_1-{\bf x}_2)} \rho ({\bf x}_1,{\bf x}_2) 
\label{8}
\end{eqnarray}
\noindent
in terms of the differential cross-section 
$W({\bf q})=\int d\omega \, W({\bf q},\omega)$. We note that such a simple
correspondence between the OPDM and the crossection is due to the scattering
length approximation of the two-particle interaction (see Eq.(\ref{2})).
If the momenta of the fast atoms are larger than $1/a$, no simple relation
can be obtained in general, and only an estimate of the mean correlation
length can be extracted from $W({\bf q})$. 

Now let us consider the {\it time-reversed} process. 
In this case the initial state consists of two particles
with momenta ${\bf k}_1$ and ${\bf k}_2$ and $N$ atoms
in the cloud. In the final state there are one fast 
atom with momentum ${\bf k}$ and $N+1$ atoms in the cloud.
We note
that the process under consideration can be viewed
as the 4-wave mixing of the matter waves (see \cite{4WAVE}
and references therein). In other words, two incoming beams
of fast and coherent atoms aimed at the cloud may produce
a third (outcoming) atomic beam.  
Its intensity $J({\bf n})$ per unit solid angle
in the direction of some unit vector ${\bf n}$
is given by the Golden rule formula where the perturbation
is the second term in Eq.(\ref{4}), and the finite state summation
should be performed over the absolute value of  ${\bf k}={\bf n}k$.
Finally, we arrived at the expression

\begin{eqnarray}
J({\bf n})={8a^2n_{{\bf k}_1}n_{{\bf k}_2}\over \pi m  }
\int \! d\omega 
k(\omega )\int \! \! \int d{\bf x}_1 d{\bf x}_2
\int_{-\infty}^{\infty} dt \,
{\rm e}^{i{\bf q}({\bf x}_1-{\bf x}_2)-i\omega t} \,
\rho^{(r)} ({\bf x}_1, t; {\bf x}_2, 0) \;
,
\label{9}
\end{eqnarray}
\noindent
where the notation
 ~$\rho^{(r)} ({\bf x}_1, t; {\bf x}_2, 0)=
\langle \psi ({\bf x}_1, t)
\psi^{\dagger}({\bf x}_2, 0)\rangle $~
is introduced; $ n_{{\bf k}_1},\! n_{{\bf k}_2}$ stand for
the atomic densities of the incoming atomic beams;
the momentum ${\bf k}= {\bf n}k(\omega )$ of the outcoming
atom carrying the energy excess $\omega$
and the momentum excess ${\bf q}$ (if compared with the total
energy and momentum
 carried by the two incoming atoms) is given by 

\begin{eqnarray}
{\bf n}k(\omega)={\bf q}+{\bf k}_1 + {\bf k}_2,\quad
k(\omega)=\sqrt{k^2_1+k^2_2+2m\omega}.
\label{10}
\end{eqnarray}
 \noindent
Note that when $r_c$ is large, ${\bf q}$ can be set to zero
in Eq.(\ref{10}). This implies, that $\omega $ 
can be effectively selected by choosing values of 
${\bf k}_1,\;  {\bf k}_2$ so that ${\bf k}_1{\bf k}_2=m\omega$
to satisfy the requirement of the energy and
momentum conservation. 
 
Consider the structure of $W({\bf q})$ and $ J({\bf n})$
in the most characteristic cases. 
In the case of a {\it pure Bose-Einstein condensate}
characterized by the condensate wave-function 
$\exp(-i\mu t)\Phi ({\bf x})$, one finds
~$\rho ({\bf x}_1,{\bf x}_2) = \Phi^*({\bf x}_1) \Phi ({\bf x}_2)$~
and ~$\rho^{(r)} ({\bf x}_1, t; {\bf x}_2, 0)=\exp (-i\mu t)
\rho ({\bf x}_2,{\bf x}_1)$~. This yields 

\begin{eqnarray}
W({\bf q})  =   8\pi a^2 | \Phi_{\bf q} |^2, \quad
J({\bf n})=2n_{{\bf k}_1}n_{{\bf k}_2} {k(\mu )\over \pi m}
W({\bf q}),
\label{11}
\end{eqnarray}
\noindent
where $\Phi_{\bf q} = \int  d{\bf x} \, {\rm e}^{-i{\bf q}{\bf x}} \, 
\Phi ({\bf x})$. 
In a trap having a center of symmetry,  $\Phi ({\bf x})$ is
characterized by a definite parity. Thus, $\Phi_{\bf q}$ can be taken real. In
this case  $\Phi ({\bf x})$ can be obtained from 
$W({\bf q})$ ( measured either in the {\it direct}
or the {\it reversed} process). 

An instructive case is the {\it axially symmetric quantum 
vortex}. The presence of a vortex carrying the vorticity
$l=\pm 1,2,3...$ in the center of 
the axially symmetric condensate drastically changes
the scattering pattern. Indeed, in this situation, 
$\Phi =\exp(il\theta )\sqrt{n(r,z)}$, where $\theta $
is the axial angle and $n(r,z)$ stands for the axially 
symmetric condensate density as a function
of the distances $r,\, z$ perpendicular to the axis
and along the axis, respectively. Accordingly,
Eq.~(\ref{11}) shows
that $W({\bf q})=0$ for ${\bf q}$ directed along the 
vortex line. The differential cross-section becomes
finite as long as there is  a component 
${\bf q}_{\perp}$ of ${\bf q}$ perpendicular
to the axis (or the vortex displaces from the condensate
center). In this case, $W({\bf q})\sim q_{\perp}^{2l}$
for $ q_{\perp}\to 0$. 

Quite similar to the quantum vortex is the case of
the {\it supercurrent state} of  toroidal Bose condensate.
The suppression of $W({\bf q})$ for ${\bf q}$ perpendicular
to the plane of the torus permits the distinguishing of the
supercurrent state from the currentless genuine ground state.

Now let us consider the possibility of measuring the 
{\it relative phase} between two condensates by the discussed
method. The problem of detecting a relative phase
has been addressed many times in different contexts
(see Refs.\cite{DALI,WALL}
and references therein). 
The Bose field inside a split trap can now be represented as
\cite{WALL}

\begin{equation}
\psi =a_1\psi_1({\bf x})+
 a_2\psi_1({\bf x -x}_0)
\label{14}
\end{equation} 
\noindent
where
$a_1$ and $a_2$ destroy an atom in the condensate
number $1$ and $2$, respectively; $\psi_1({\bf x})$
stands for the single-particle state of the first
condensate; the single-particle state of the
second condensate $\psi_1({\bf x -x}_0)$ is
displaced by ${\bf x}_0$ from the position 
of the first one. We assume that
both states have no spatial overlap.
Making use of the ansatz (\ref{14})
in Eq.(\ref{8}), one finds

\begin{equation}
\displaystyle W({\bf q})=
8\pi a^2|\psi_{1{\bf q}}|^2
[\langle a^{\dagger}_1a_1
\rangle +\langle a^{\dagger}_2a_2\rangle 
+(\langle a^{\dagger}_1a_2\rangle 
{\rm e}^{-i{\bf qx}_0} + h.c.)]
\label{15}
\end{equation}
\noindent
where ~$\psi_{1{\bf q}}$~ stands for
the Fourier transform of $\psi_1({\bf x})$.
The means 
$\langle a^{\dagger}_1a_1\rangle =N_1$
and $\langle a^{\dagger}_2a_2\rangle =N_2$
represent the mean numbers $N_1,\,N_2$
of atoms in the condensates. 
The interference
effect is described by the
crossterm $\langle a^{\dagger}_1a_2\rangle$.
If these two
condensates are in a coherent superposition,
a well defined relative phase 
$\varphi$ can be introduced, so that
$\langle a^{\dagger}_1a_2\rangle =
\exp (i\varphi )\sqrt{N_1N_2}$.
Accordingly, Eq.(\ref{11})
acquires a form

\begin{equation}
\displaystyle W({\bf q})=2 W_1({\bf q})(1+ \cos(\varphi -{\bf
qx}_0)),
\label{16}
\end{equation}
\noindent
in the case $N_1=N_2=N$.
Here $ W_1({\bf q})=8\pi a^2|\psi_{1{\bf q}}|^2N$
 stands for the distribution 
which would be produced if only one condensate were
present. Eq.(\ref{16}) implies an extremely sharp dependence
of $W({\bf q})$ and of $J({\bf n})$ in Eq.(\ref{11})
on $\varphi$. 

In the situation when the two 
condensates were initially in the number states  
$|N_1,N_2 \rangle$, no relative
phase existed before the measurement. This
implies that initially 
$\langle a^{\dagger}_1a_2\rangle =0$
in Eq.(\ref{15}). Then,
in the course of the scattering and {\it measuring}, the
relative phase should be built
in accordance with the general understanding
\cite{DALI,WALL}. In
order to describe the dynamics of the phase
formation, the evolution of the 
correlator $\langle a^{\dagger}_1a_2\rangle$
should be found, where $\langle ...\rangle$
is understood as the conditional quantum 
mechanical mean under the condition that
a certain amount of the scattering
events has been detected. 

In the absence of a genuine condensate (or for the above-the-condensate
part of OPDM) one normally deals with the so-called 
{\it quasi-homogeneous regime}, when the off-diagonal correlation 
radius $r_c$ is much less then 
the scale of density variation given by, e.g.,
the size $R_c$ of the atomic cloud. 
In this case, it is reasonable to introduce the variables
~${\bf r} = {\bf x}_2 - {\bf x}_1$~ and 
~${\bf R} = ({\bf x}_1 + {\bf x}_2)/2$, and to represent
$W({\bf q})$ as 
\begin{equation}
\displaystyle  W({\bf q})= 8\pi a^2 \int d {\bf R} \, 
\rho_{\bf q}({\bf R})  \; , \; \; \;  
\rho_{\bf q}({\bf R}) = \int  d{\bf r} \,  {\rm e}^{-i{\bf q}{\bf r}} \,
\rho({\bf r},{\bf R}) \; . 
\label{17}
\end{equation}
\noindent
In the quasi-homogeneous regime, the OPDM in the Wigner representation
($\rho_{\bf q}({\bf R})$) has the  semiclassical meaning of local
(at the point ${\bf R}$) distribution of the particle momentum ${\bf q}$.
Such interpretation is justified 
by the condition $r_c \ll R_c$ and by the
definition of $r_c$ as a typical distance over which 
$\rho({\bf r},{\bf R})$ decays as a function of $r$.
We thus see that in this case $W({\bf q})$ yields spatially
averaged momentum distribution. Moreover, without contradiction
with the uncertainty principle, this averaging can be 
partially (totally) removed by collimating the incident (both
the incident and outcoming) beams. Even without removing
the averaging, $W({\bf q})$ contains valuable information about
long-range correlations in the system, since the averaging 
does not affect the order-of-magnitude value of the correlation
radius $r_c$.
If the system contains both the condensate and the quasi-homogeneous
above-the-condensate fraction, the cross-section $W({\bf q})$
is given by a combination of Eq.(\ref{11}) and Eq.(\ref{17}).

Let us discuss general requirements for the atomic detector and 
the atomic sources.
 Currently, the only realistic
detecting scheme we envision is that suggested in
Ref.\cite{ANDREWS}. This method relies on imaging the atomic
interference fringes by resonant light \cite{COHER}.
In our case, a typical angular distance between the fringes
is given by $\alpha \approx 1/r_ck \ll 1$ 
as long as the conditions (\ref{1})
and $r_c \gg n^{-1/3}$ are fulfilled. Choosing $n=10^{15}$cm$^{-3}$,
$a=3\cdot10^{-7}$cm, one finds the gas parameter 
$\xi\approx 3\cdot 10^{-5}$. Then, in accordance with
Eq.(\ref{1}), $k\approx 2\cdot 10^4 - 3\cdot 10^6$cm$^{-1}$
and for $r_c\approx 10^{-3}$cm the angular resolution
should be $\alpha \approx 5\cdot 10^{-2} - 5\cdot 10^{-4}$.
The presence of the trapping field characterized
by the change $U$ of the potential energy will distort the fringes
by producing the uncertainty $\delta k \sim m|U|/k$ of the momentum. 
This introduces
the limitation $\delta k < 1/r_c$ or
 $|U|/\mu < ka^2/(\xi r_c)$. For the chosen parameters,
it implies $|U|/\mu \leq 1$. 
A source of fast \cite{CANNON} and coherent atoms 
-- the atomic laser \cite{LASER} -- must have
an uncertainty $\Delta k$ of
the incoming momenta less than
$1/r_c$.
Thus, 
$\Delta k /k = \alpha $. Recently 
this ratio has been achieved as small as $0.09$ \cite{PHILL}.   
We note, however, that this is still not enough for
resolving $r_c$ comparable with typical condensate sizes.
Nevertheless, detecting the 
initial stages of the ODLRO formation appears to be quite feasible 
because $r_c$ in 
this case can be as small as $\sim 10^{-5}$cm, which
relieves the restriction on $\Delta k /k $ as well as that 
on $|U|/\mu$ by a factor of $10^2$.

In conclusion, we have suggested a method of scattering
fast atoms in a pure enough momentum state 
off a trapped atomic cloud in order to
test directly the one-particle density matrix of this
cloud. The differential cross-section
of the inelastic process, when one incoming fast
atom produces two fast ones, as well as of the reversed one
allow measuring
the correlation length of the local off-diagonal order. 
This gives, in particular, a powerful tool for testing 
different scenarios for the formation of the off-diagonal 
long-range order in the traps. 
This method can also be employed for detecting quantum
vortices and supercurrent states, 
as well as the effect of quantum depletion
of the condensate. A relative phase of the split-condensate
can be measured as well. 

We are grateful to M. Andrews for useful discussions of 
means of practical realization of the proposed
method.
A.B.K acknowledges a support from the PSC-CUNY Research Program.
 B.V.S. acknowledges a support from the Russian Foundation for Basic
Research (Grant No. 98-02-16262) and from the European Community (Grant 
INTAS-97-0972).


\begin{references}
\bibitem{PAIR}
The identification of pairs resulted from one and the
same scattering event can be based on the fact that
the angle between
the two created fast atoms is always close to $\pi /2$
as long as $r_ck\gg 1$.
\end{references}
\end{document}